# Characterizing Pairs Collaboration in a Mobile-equipped Shared-Wall Display Supported Collaborative Setup

**Razan N. Jaber** (*razan@dsv.su.se,* Department of Computer Science, Stockholm University, Sweden)

**Ragaad AlTarawneh** (*ragaad.altarawneh@intel.com*, Intel Research Labs, USA)

**Shah Rukh Humayoun** (*humayoun@cs.tufst.edu*, Department of Computer Science, Tufts University, USA)

**Abstract**   Recent advancements in mobile devices encourage researchers to utilize them in collaborative environments as a medium to interact with large shared wall-displays. In this paper, we focus on a semi-controlled user study that we conducted to measure the collaborative coupling ratio between partners working in pairs in a collaborative setup equipped with a shared tiled-wall display and multiple mobile devices. We invited 36 participants in 18 pairs to take part in our experiment in order to analyze how they communicate and collaborate with each other during the experiment. We observed their collaborative coupling by measuring how often they verbally and visually communicated. Further, we found frequently used collaborative physical position patterns by observing the pairs' physical arrangements and standing positions. Moreover, we combined these factors to gain a clearer understanding of coupling in our setup, taking into account the mobility factor offered by the mobile devices.  Results of the study show interesting findings about the coupling factors between the partners mainly due to the flexibility offered by including mobile devices in our collaborative setup.

**Keywords**   Computer-Supported Cooperative Work (CSCW); Tiled-Wall Collaborative Environment; Coupling; Smart Devices; User Study; Collaborative Relation Building.

## 1.   Introduction

Nowadays we witness high usages of smart mobile devices (e.g., smart watches, smartphones, tablets, etc.) for performing many activities of our daily life, e.g., receiving calls to monitoring health fitness. This is mainly due to the advancements that have been made in this domain during the last decade. These advancements enable us to use mobile devices intuitively and interact with them freely without extra efforts. At the same time, interactive collaborative systems are growing in popularity due to their important role in group working environments (Babuška and Groen, 2010). Many previous studies (e.g., Guimbretière et al. 2001; Tang 1991) have shown the advantages of these emerging technologies to enhance the quality and the quantity of generating ideas in teamwork environments. This holds especially for the emerging technology of collaborative systems (e.g., multi-touch large displays, tiled-wall displays, interactive tabletops, etc.), which are expected to change teamwork environments in many application domains. Therefore, collaborations through computer-supported environments have become increasingly critical in the modern world, as several members from different departments can work, interact, share ideas, and provide solutions together more efficiently and effectively.

Different settings have been proposed by researchers with the common concept of a shared screen to perform some common tasks collaboratively (a few examples are: AlTarawneh et al. 2014; Baldauf et al. 2012; Haller et al. 2005; Heinrich et al. 2014; Humayoun et al. 2015; Isenberg et al. 2010; Isenberg et al. 2012; Jakobsen and Hornbæk 2014; Jones et al. 2012). Amongst them, two kinds of environments are the most common ones: tabletop collaborative environments (e.g., Haller et al. 2005; Heinrich et al. 2014; Humayoun et al. 2015; Isenberg et al. 2010; Isenberg et al. 2012; Jones et al. 2012), where users stand or sit around the tabletop display and perform collaborative tasks; and *wall-display collaborative environments* (e.g., AlTarawneh et al. 2014; Baldauf et al. 2012; Jakobsen and Hornbæk 2014), where users stand in front of a large shared wall-display and perform collaborative tasks. In the case of tabletop collaborative environments, limitations that influence users in building collaborative relations could be the interface orientation on the tabletop, leading roles between users, and space limitations (Hall 1988; Scott et al. 2004). These limitations could be reduced in wall-display collaborative environments as they give freedom to move, equal role in collaboration, and a common angle to watch the screen. However, most of the wall-display collaborative setups are built either with a multi-touch wall-display, e.g., setup by Jakobsen and Hornbæk (2014), or a mouse controlled wall-display, e.g., setup by Birnholtz et al. (2007)). These setups have their own limitations, e.g.: some restrictions in freedom of movement, single interaction with an object at one time, unintentionally hiding some parts of the screen when one user touches the screen, etc. In most of the previous studies for collaborative environments (e.g., Isenberg et al. 2010; Jakobsen and Hornbæk 2014; Tang et al. 2006) users' positions were mostly located around (i.e., in the case of tabletop-supported setups) or near (i.e., in the case of wall-display supported setups) the shared display.

Recently, researchers have also focused on utilizing smart mobile devices (i.e., smart watches, smartphones, and tablets) to interact with large shared displays, e.g.: Baldauf et al. (2012) built a setup where users worked with smartphones to interact with a wall-display and conducted a user study on a public screen to find out the general acceptance towards such a setup; AlTarawneh et al. (2014) provided a setup where multiple users can interact simultaneously with a shared tiled-wall display with smartphones and tablets to perform tasks collaboratively. Such setups have multiple benefits over the previous ones like users' freedom to move around the environment and changing their positions, full view of the screen, and simultaneous interaction with the application interface. These kinds of setups (which use smart mobile devices for interaction) can be utilized more effectively and efficiently from the collaboration perspective, e.g., the proposed setup by AlTarawneh et al. (2014).

In mobile devices equipped collaborative setups, we need to measure how such setups affect building effective collaborative relations (i.e., the *collaborative coupling*) between the partners working in these setups. Many factors in these setups (e.g., the freedom of mobility or the overall view of the wall without any blocking) could impact the strength of building a collaboration between the partners. However, building a successful collaborative relation is controlled mostly by the verbal communication, the visual communication, and in the mentioned setups the physical alignment of team partners (Heinrich et al. 2014; Jakobsen and Hornbæk 2014). All these factors together lead to *tightly collaborative coupling*, which could indicate the strength level of the collaborative relation.

The addition of mobile devices for interaction in a shared collaborative space equipped with a large wall-display provides more spatial and collaborative flexibility for team members. Therefore, we aim to investigate how the usage of mobile devices in a shared wall-display collaborative environment assists the team members to execute their tasks in a more collaborative fashion. Therefore, we raised the following research questions in this work:

- How does the usage of the mobile devices as an interaction medium in a shared wall-display collaborative setup impact the communications between the team members?
- Does using mobile devices as an interaction medium in a shared wall-display collaborative setup impact the group awareness, the participants' coupling style, and the participants' satisfaction level towards the underlying setup?

To answer these raised questions, we conducted a laboratory-controlled user study with 18 pairs in a mobile devices equipped collaborative setup, which we built mimicking the setup by AlTarawneh et al. (2014), i.e., a large shared tiled-wall display and multiple mobile devices for interaction. The main focus of this paper is on the conducted user study and its findings to measure the *collaborative coupling* *factors* between the pairs (i.e., in total 18 pairs) who worked in this collaborative setup. We videotaped and audio recorded all sessions, then we analyzed the users' interactions in terms of different factors. Our objective was to observe users' verbal communication and visual communication levels. Due to the mobility that was offered by the mobile devices in the built collaborative setup, we observed the pairs' physical positions during the test. Then we correlated it with users' verbal and visual communications in order to get an indication of the coupling ratio. Results of the study show interesting findings about the coupling factors between the partners mainly due to the flexibility offered by including the mobile devices in our collaborative setup.

The remainder of the paper is structured as follows: In Section 2, we provide background on collaborative coupling and descriptions of other collaborative setups. In Section 3, we provide details of our built collaborative setup. In Section 4, we explain the collaborative model that we used in our study. In Section 5, we describe details of our conducted user study. In Section 6, we focus on the results and the findings of our study. Finally, we conclude the paper in Section 6.

## 2. Background and Related Work

### 2.1 Collaborative Coupling

Gutwin and Greenbergs (2002, p. 20) defined a general term of coupling which describes it as "the amount of work that one person can do before they require discussion, instruction, action, information, or consultation with another person". According to them, team members transfer from a loose to a tight coupling condition when they see the opportunity to make the decision for their next activity or when they require the other persons to collaborate with them on the current task. Therefore, it has been concluded from previous work that people in a collaborative environment move between individual working conditions (that can be referred as loose coupling conditions) to shared working conditions (that can be referred as tight coupling conditions) (Baker et al. 2013; Dourish et al. 2006; Gaver 1991; Heath et al. 1994; Salvador et al. 1996).

Olson and Teasly defined task-coupling as a classification of three categories: tight, moderate, and loose based on "how immediate a response is needed, and how much interaction is required for either clarification or persuasion" Olson and Teasly (1996, p. 422). A simple definition of a collaborative coupling is also provided by Pinelle et al. (2003, p.302): "coupling in collaboration is the degree to which people can work as individuals before needing to interact with another member of the group". According to them, group members are considered in a loosely coupled collaboration if they can perform more individual tasks in the environment, compared to a tightly coupled collaboration where a close collaboration is required to perform tasks; thus, reducing the individual workspace. Further, they described that the coupling factors in a collaborative environment heavily depend not only on the task (as it defines how much close collaborations or interactions amongst the team members are required) but also on the group itself, as many groups prefer to work together compared to ones that do not. Therefore, tightly coupled interactions may introduce interdependencies between the team members in a group compared to the loosely coupled interactions.

Another important term in collaborative work is *mixed-focus collaboration*, which can be defined as the continuous switching between "individual" and "group" work, i.e., the frequent transition between *loose coupling* and *tight coupling* (Dewan et al. 2010). However, it is important to distinguish the mixed-focus collaboration from the multi-view collaboration, which allows different users to work independently on different aspects of the shared objects without allowing them to be aware of the view or the actions of the other collaborators. Therefore, previous researchers have suggested workspace awareness mechanisms in order to allow users to work independently while being aware of aspects of others' views and actions. The *workspace awareness* is an important factor in collaborative environments that affects the coupling levels, which is described as involving personal understanding of interacting with the shared workspace (Gutwin and Greenbergs 1996; Gutwin and Greenbergs 1998; Gutwin and Greenbergs 2002; Shupp et al. 2007). The awareness mechanisms gain special importance in the case of mixed-focus collaborations, as users frequently switch between the "individual" and the "group" working modes (Dewan et al., 2009).

Tang et al. (2006) related the coupling with *mixed-focus collaboration*, which is used to describe tasks that require switching between an independent and a shared activity, and *workspace awareness* (Gutwin and Greenbergs 1996; Gutwin and Greenbergs 2002). However, they referred to collaborative coupling in their collaborative environment setup (i.e., over a tabletop) as "*the manner in which collaborators are involved and occupied with each other's work*" Tang et al. (2006, p. 1181). Therefore, the focus was on the team members' linked activities to judge the coupling level (i.e., through mixed-focus collaboration) and on the task semantics (Pinelle and Gutwin, 2005).

In an evaluation study, Dewan et al. (2009) evaluated the side-by-side software development approach in which programming pairs worked on a single task while they were allowed to work concurrently on two different computers and were able to see each other's display. The authors considered this kind of collaborative programming approach as mixed-focus collaboration, as it requires a frequent transition between the individual working mode and the group working mode. Later, Dewan et al. (2010) proposed the coupled tele-desktops approach, which is a two-person interaction mechanism that is meant to not be biased towards individual or group work. They evaluated the mixed-focus concept and the proposed tele-desktops approach using a quantitative framework consisting of two main components: a set of precisely-defined coupling modes determining the extent of individual and group work, and the time spent to change between these modes.

In our work, we use the definition of the collaborative coupling as a combination of *mixed-focus collaboration* and *workspace awareness*, due to the usage of mobile devices as an interaction medium with the shared wall-display in our collaborative setup (that we explain in Section 3).

## 2.2 Collaborative Setups with Shared Wall-Display

Many previous studies have been conducted in the past to show the feasibility and effectiveness of shared large wall-displays in collaborative environments to understand users' behavior and collaboration in such environments. In one of the earlier studies done by Elrod et al. (1992), the authors provided their general observations about the interaction with large displays used in office environments and explained how users can organize their activities in such environments. While Guimbretière et al. (2001) used brainstorming sessions with five groups of designers for evaluating a high-resolution wall-display, and concluded that users reacted positively to the wall interaction metaphor; however, they did not describe how these groups worked together around the display.

Later, many studies were conducted using large wall-displays to show improvements in performance and user satisfaction for many tasks, including personal desktop computing (e.g.: Ball and North 2005; Bi and Balakrishnan 2009; Czerwinski et al. 2003), sense making (Andrws et al., 2010), map navigation (Ball and North, 2005), 3D navigation (Desney et al., 2006), etc. One of the main advantages of large wall-displays is that users can view multiple windows at the same time with reduced navigation efforts. Several studies have shown this fact, e.g., Bi and Balakrishnan (2009) studied users working with a 5 meters display and reported that a large wall-display can improve users' awareness of peripheral applications. Ball and North (2005) conducted an observational study and interviewed users working with a wall-display consisting of 3x3 LCD panels. They showed that it helped users to switch between the tasks and enhanced their awareness for the secondary tasks. They compared simple navigation tasks using a zoom-pan interface with one, four, and nine tiled monitors. They found out that participants performed the given tasks faster and felt less frustrated when they worked with nine monitors compared to working with other setups.

Many studies have also been conducted for investigating group behavior in wall-display supported environments. Few examples are: Huang et al. (2006) identified common factors that influence groupware when using a large wall-display. Birnholtz et al. (2007) performed an exploratory study for checking the group behavior and performance in collaborative tasks with a wall-display setup through two interaction settings, a single shared mouse device and one mouse device per user. They found out that using a shared mouse device triggered more discussion while using multiple mouse devices allowed more parallel work amongst the users. Hawkey et al. (2005) used a display equipped with a SmartBoard touch sensitive overlay, which allowed annotations to be drawn on the top of the map. They were interested in finding out the effect of using the physical proximity to the shared display and the usage of an input device on users' collaboration. They found out that the proximity between the partners and their distance from the display affect the collocated collaboration between the partners.

Recently, Jakobsen and HornbÆk (2014) conducted a user study in which they analyzed users' behaviors, use of gestures, body orientation, and users' positions while working on a high-resolution multi-touch wall-display. They focused on how pairs of users physically navigate information on the wall-display and tracked the participants' positions to quantitatively describe their movements in front of the display. In order to evaluate the system, they collected participants' answers using questionnaires and taking notes of their comments during debriefing about their use of the display. They defined more physically based codes, which are more easily identified during coding compared to more socially based codes. Baldauf et al. (2012) conducted an evaluation study for detached multiuser interaction with large wall-displays. In their study, smart phones were used to interact with the large displays for controlling interactive applications. They conducted a user study at a public science event to collect feedback, observed users' behaviors and took notes during the usage of the system. The main goal of their study was to find out the general acceptance and if the participants would be prepared to use the system.

2.3   Collaborative Setups with Shared Tabletop Display

Many studies have also been conducted focusing on tabletop collaborative setups. A few examples are: Haller et al. (2005) presented an evaluation study of "Coeno", which is a computer enhanced face-to-face presentation environment for discussions using tabletop technology in combination with digital information. Buisine et al. (2007) investigated the usability and the usefulness of interactive tabletop technologies in supporting group creativity. In their setup, the tabletop interface enabled groups of four participants to build mind-maps. In their study, they compared the use of a tabletop to the traditional paper-and-pencil mind-mapping sessions. Maldonado et al. (2010) conducted a study to evaluate a multi-touch tabletop setup called "Cmate", which can be used for collaborative concept mapping. They evaluated the usability of their setup through a qualitative formative evaluation approach. Jones et al. (2012) presented a collaborative work environment, consisting of an interactive tabletop and an interactive board, in order to explore phases beyond the brainstorming in preliminary design. Heinrich et al. (2014) evaluated users' behavior while sitting in different orientations around a tabletop during financial advisory sessions. They used the prototype in a realistic setting with four experienced financial advisors and conducted two experimental studies, one with traditional pen and paper advisory session and the other one with the IT-supported advisory.

Christophe et al. (2014) conducted studies using augmented-view tabletop systems through collaborative map-based tasks. They found improvements in the group's performance when the view configuration matches the spatial arrangement of the group's tasks. Wallace et al. (2013) investigated users working on collaborative sense-making tasks using three setups: only tabletop, only tablets, and with both. They found improvements in sense making performance using tabletop mainly due to the shared workspace, as it supports the group in prioritizing the information or to do comparisons between the task data. However, they found a negative correlation in the quality of team members' participation when it comes to using the tablets.

Recently, Zagermann et al. (2016) studied the effect of the size of a shared horizontal interactive workspace on users' attention, awareness, and efficiency during the cross-devices collaboration. They conducted their study using 15 pairs executing a sense-making task with two personal tablets and a horizontal shared display of varying sizes. Their findings show that different display sizes lead to differences in participants' interactions with the tabletop and in the groups' communication styles. However, they found that larger tabletops do not necessarily improve collaboration or sense-making results, because such displays can divert users' attention away from their collaboration. They concluded that the collaboration between the group members should depend on the task at hand rather than the size of the shared screen, as larger tabletop display size may not be a better choice.

## 3. The Environment Setup

We built our collaborative environment setup (see Figure 1 and Figure 2) mimicking the setup initially developed by AlTarawneh et al. (2014), which consists of a large tiled-wall display and multiple smart mobile devices for the interaction. The tiled-wall in our case was a vertical display containing nine monitors (in a 3x3 matrix with a total tiled-wall display resolution of 7680 X 4800, a total of 36,864,000 pixels) with a height of 78 inches and a width of 47 inches. The room dimensions were 5.5 meters by 5.3 meters (see Figure 2), which gives the facility to users to move freely around the environment to choose their preferred standing position and their style of interaction during the collaborative tasks.

The built setup was scalable in the sense that it supported simultaneously multi-users' interactions using a mobile app on their mobile devices, which in our study were Apple iPad devices in our study, through with WIFI connections (see Figure 1 for an abstract representation). The mobile app on the mobile devices used for interacting with the application on the tiled-wall provided a menu-bar for eye-free interaction, same as proposed by Bauer et al. (2013). Further, it provided a list of options to interact with the application as per requirements of the targeted goals.

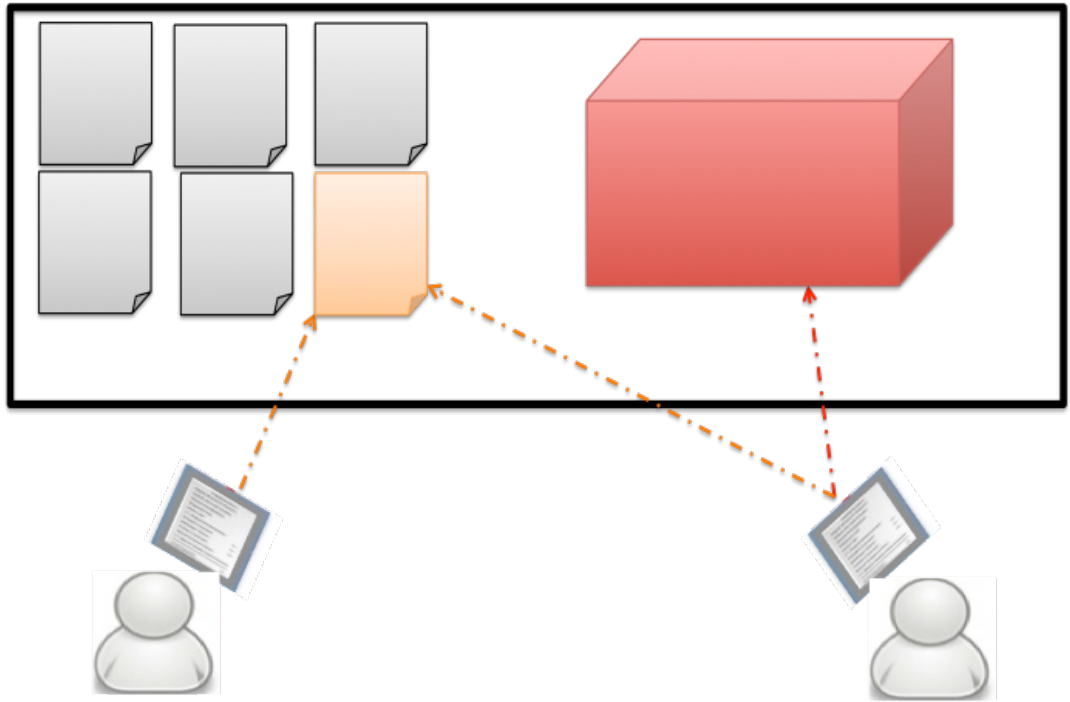

**Fig. 1** Abstract representation of the built collaborative setup. Dash-arrows show that users can interact with different artifacts/views as well as with the same artifact/view simultaneously through their mobile devices.

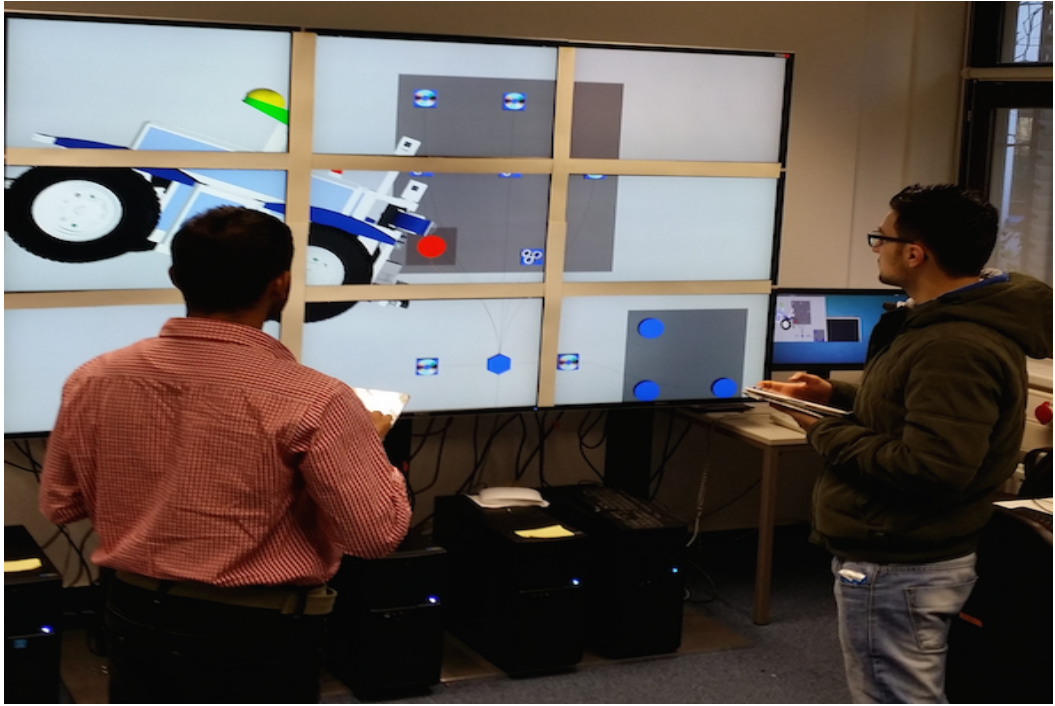

**Fig. 2** The built collaborative setup in the actual form.

In our study, we used the ESSAVis application (AlTarawneh et al., 2014) to perform the collaborative tasks. ESSAVis was developed to facilitate the safety analysis process in embedded systems and provides support to work with it individually as well as in a collaborative fashion. ESSAVis provides a 2D*plus*3D visual environment, where the 2D view provides the abstract graph representation of the safety scenario in the underlying system while the 3D view provides a 3D model of the underlying system. In our built setup, users (2 or more) were able to interact with different artifacts/models or with the same artifact/model in one or both views simultaneously. Therefore, it supported the tasks that were needed to be performed in the *loosely coupled* fashion as well as the ones that were needed to be performed in the *tightly coupled* fashion. Compared to the other setups mentioned earlier, our built setup provided the users the facility to simultaneously interact with the same artifact/model or multiple artifacts/models through their mobile devices as well as to move freely around the environment without blocking the view of other group members.

## 4. The Collaborative Model

The abstract representation of our built setup, shown in Figure 1, illustrates the idea of having multiple users interacting with the same or different artifacts/models and with identical or different views of the used ESSAVis application, rendered on the shared tiled-wall display. However, we observed two types of communications in our built setup: the first one was between the partners themselves, while the second one was between the users and the system.

- **Communication between the partners:** This indicates any type of interaction between the partners working together before or during execution of the task. In general, this interaction impacts the quality of building a collaborative coupling between the partners. This interaction includes two forms of communication between the partners, the *visual* and the *verbal*:
    - ***The Visual Communication:*** It is based on how often the partners have any kind of visual communication. In our study, we classified it into two categories:
        - *Glancing:* When one partner pays attention to the other partner in certain situations. Glancing is important for the partners to maintain awareness of what the other one is doing, and for coordinating their work (e.g., indication for switching to the next task).
        - *Looking:* When one partner gives at least a 5 seconds long gaze (this time duration was proposed by Jakobsen and Hornbæk (2014)) to the other partner in order to indicate some meaning (e.g., to signify that the task has been completed or that now it is the other partner's turn to continue the work).
    - ***The Verbal Communication:*** It was defined based on how often the partners talked with each other. Talking can be divided into *ordinary conversation* and *verbal shadowing*, where people are engaged in a conversational sequence, or talking to no one in particular but for all to overhear. Both types are used to help people understand what the other partner is doing (Pinelle and Gutwin, 2005). We measured it using:
        - *Talking Factor:* When one partner or both partners talk for at least 5 seconds (e.g., to collaborate during the task or to talk about the next task).
    - The Visual+Verbal Communication:
        - *Talking & Looking:* When one partner or both partners talk and give a long gaze for at least 5 seconds. We observed that in the case of less than 5 seconds visual+verbal communication, partners of the pair just said one or two words to acknowledge (e.g., *OK* or *yes*); therefore, we made it part of the *glancing* category rather than defining a new one.

- **Communication between users and the system:** This indicates the number of steps that are required from any user to complete a particular task. This measurement is useful in judging the usability of the underlying application. However, the scope of this paper is the relationship between the partners in the underlying collaborative setup rather than measuring the usability of the application running on the setup.

To measure the quality of the collaborative coupling between the partners, we focus on measuring the *coupling ratio* that indicates the factor of how the partners verbally and visually are connected to each other (Heinrich et al. 2014; Jakobsen and Hornbæk 2014; Kleinke 1986). As discussed earlier and also described by Jakobsen and Hornbæk (2014), those who have higher levels of verbal and visual communication in the collaborative tasks can be considered to be more *tightly coupled* pairs. Moreover, in our setup users have more freedom to move around the environment due to the usage of mobile devices for interacting with the tiled-wall display compared to other setups where users are required to interact by directly touching the touch-wall display (e.g., Jakobsen and Hornbæk, 2014) or the tabletop surface (e.g., Heinrich et al. 2014; Isenberg et al. 2010; Tang et al. 2006). Therefore, in order to measure the *collaborative coupling level* between the partners, we need to take care of their physical positions aside from their verbal and visual communications. In other words, we use the definition of partners' coupling by including the physical alignment of the partners (i.e., the *workspace awareness*), as it impacts the verbal and the visual connectivity between the partners. This is because partners in our collaborative setup may have strong verbal and visual communications if they are physically standing in a way that would help them in collaborating together and minimizing the efforts required following the changes that take place on the shared tiled-wall display.

## 5. The User Evaluation Study

We conducted a controlled observational user study in a laboratory setting with 18 pairs, who worked on our built setup with a predefined list of tasks that required close collaborations amongst the participants in pairs. Like Isenberg et al. (2012) and Zagermann et al. (2016), we chose a group size of two because in these studies two group members working on collaborative tasks were already sufficient to identify differences in coupling styles and in investigating the collaborative environment. In the remainder of this section, we provide design details of our study.

## 5.1 Goals and Research Questions

The goal of our controlled user study was to measure the influence of using smart mobile devices as interaction devices on users' collaborative coupling in collaborative setups with a large shared wall-display.

To achieve this goal, we conducted the controlled study to explore users' behaviors working around a shared wall-display setup (as described earlier in Section 3) while they were using tablet devices to interact with the system in order to achieve their tasks. We measured the talking and the looking ratio of each pair while they were freely moving around the shared wall-display. We wanted to determine what effects the freedom, offered by mobile devices during interaction with the shared large wall-display in a collaborative environment, had on improving the collaborative coupling level between the pairs in performing the collaborative tasks. Based on this, we specified the following research questions for our study:

- Research Question 1 (RQ1):
    - How does the usage of the mobile devices as an interaction medium in a shared wall-display collaborative setup impact the communication between the team members?
- Research Question 2 (RQ2):
    - Does using mobile devices as an interaction medium in a shared wall-display collaborative setup impact the group awareness, the participants' coupling styles, and the participants' satisfaction levels towards the underlying setup?

## 5.2 Study Design and Procedure

Based on the identified goal and the raised questions, we designed the study as a controlled observational study under laboratory conditions with 36 participants. These participants worked in pairs (resulting in 18 pairs) to perform the predefined list of tasks in our collaborative setup with an approximate time frame of one hour per session for each pair.

The participants were chosen randomly amongst the 80 students from a semester HCI course, a mixture of master and bachelor level students with a background either in Computer Science, Social Informatics, or Cognitive Science. Out of 36 participants, 9 were females. Partners in 13 pairs knew each other well, partners in 1 pair knew each other as classmate, while partners in the remaining 4 pairs did not know each other before the experiment. The average age of participants ranged from 20 to 32 years with a median of 24.5 years. They were from six different countries and 26 of them were international students. Most of the participants (i.e., 68%) commented that they can do simple interaction while working on mobile devices (i.e., replying a person or communicating with a person), 11% commented that they cannot do any kind of interaction while working on mobile devices, while the remaining 16% responded with the answer "may be". All participants had considerable experience using computer interfaces due to their background; however, 92% did not have any prior experience working in any collaborative setup. It was made clear to the students that the experiment would not be graded. Further, none of the instructors or professors were present and the involved trainer was new to them. The participants' interaction with the system and with each other during the test was audio and video recorded for the later analysis.

We ensured participants' privacy by not associating their personal preferences in public reports. Video and audio recordings remain confidential, as they contain personal data about the participants' behaviors. We explained to participants the general goal of the study and asked their permission for the audio and video recordings before the start of experiment. We assured them that their personal data would not be published.

A pre-questionnaire form was used before starting each test to find out participants' background (i.e., age, gender, field of study, their relation, and their experience with collaborative setups). Afterwards a tutorial session was conducted to introduce the used ESSAVis application and the environment settings. Then 10 minutes were given to the pair to try out the whole setup. They were allowed to ask further questions during the training session and the test session in case of any technical issues. After the training session, participants of each pair were briefed about the tasks on a printed document. Participants were then asked to perform the test with the given set of tasks. After the test, each participant was asked to complete two questionnaire forms: a closed-ended questionnaire form where each question offered six different options based on a Likert scale (scaled from 1 to 5 to show the degree of agreement with each question from the participant's perspective and a sixth option "Don't know"), and an open-ended questionnaire form with general questions.

Participants were allowed to freely choose any physical arrangements and standing positions that suited them, as no instructions were given for particular arrangements or positions. They were asked to act naturally around the tiled-wall display. The participants were also asked to verbalize their thoughts and to talk freely while performing the given tasks, as well as to comment on any problems they face while using the setup. The same trainer monitored the experiment in order to avoid any differences or biases.

## 5.3 Study Tasks

The used application aims at enhancing the safety analysis process of embedded systems. This requires a close collaboration between different engineers who work together to understand the underlying embedded system structure. Therefore, we designed the test with the help of a domain expert who helped us in writing the scenarios that require close pair-wise collaboration between the partners. Our used application ESSAVis provides two side-by-side visualization views of the underlying embedded system (see Figure 2), i.e., the hardware model of the system in 3D form and the failure scenario of the system using the compound graph representation.

In total, we designed 4 tasks with different levels of difficulty where each task needed a number of steps to reach completion. Except for Task 1, tasks were given in a random fashion to reduce learning effects. Here, we provide descriptions of these four tasks:

- **Task 1 – Navigation:** In this task, both partners (i.e., User-A and User-B) want to interact with the virtual views for performing their first interaction with the tiled-wall display together. Further, they want to get used to the interaction speed with the system.
    - o **Task 1.1 – Translating the virtual views:** In this subtask, each partner works on different virtual visual views in parallel; however, both partners perform the same interaction simultaneously. User-A is interested in translating the system 3D view in four directions (i.e., up, down, left, and right), while User-B is interested in translating the abstract representation view in four directions. This task helps both partners in getting used to the system speed and the collaborative working environment.
    - o **Task 1.2: Translating the same view in opposite directions:** In this subtask, User-A is interested in translating the abstract representation view upward. Therefore, she/he expands the component in the lower part of the graph view. While User-B is interested in translating the same view downward; therefore she/he translates the 3D view to the upper part of the display and zooms it out. In this case, both partners need an effective coordination so that each one of them can achieve the desired goal.
    - o **Task 1.3 – Interacting with the same view but different actions:** In this subtask, User-A wants to rotate the system 3D model to a desired level while User-B wants to translate the same model to the center of the display.

- **Task 2 – Changing the scene configuration (conflict situation):** Both partners (i.e., User-A and User-B) want to change the configuration of the whole virtual scene based on their priority. In this case, User-A is interested in presenting the scene as a big graph view and a small 3D model view. On the other hand, User-B wants to present both views as the side-by-side layout (i.e., sharing the same screen sizes). Each partner wants to see different layout; hence, they need to interact and discuss with each other in order to be able to change the layout of the scene.
- **Task 3 – Querying safety aspects visually:** The aim of this task is to check if both partners can collaborate effectively while working on the same view in order to find out the failure path between different components in a joint work fashion. User-A wants to check the failure path between the two components, which are component numbers 10 and 11 respectively. At the same time, User-B is trying to find out the failure path between the component number 13 (which is the Front Pan Scanner component in the underlying embedded system) and the component number 14 (which is Front Slow Down component in the underlying embedded system). The ESSAVis application allows both users to simultaneously expand different components for extracting the required safety information about the underlying system, which is helpful for the partners to discuss the possible solutions.
- **Task 4 – Changing the abstract representation view layout:** This task aims at observing how the partners collaborate when they need to perform the same action on the same view. In this case, the partners need an effective interaction and collaboration so each one can reach the desired goal. In this case, User-A wants to present the abstract representation view layout in the orthogonal layout fashion, while User-B is interested in presenting the view in the graph layout circular fashion.

5.4 Participants' Behaviors Analysis Procedure

Each session was video and audio recorded in order to analyze participants' work group behavior. Pairs were allowed to ask the trainer about any confusion or difficulty they faced while executing the given tasks. Also, they were allowed to verbalize their thoughts, talk together, and help each other. We started the video analysis by watching the recordings all the way through and writing a high-level narrative of what had happened, i.e., noting down where in the video there were any potentially interesting events. We placed our focus on participants' communication behavior, their interaction with devices, as well as their physical orientation during task execution.

*Refining the video recordings*: After collecting the required field notes, video and audio recordings, we analyzed them using a two-pass open coding approach as was used by different researchers in the past (Isenberg et al. 2010; Kruger et al. 1986; Scott et al. 2004; Tang 1991; Tang et al. 2006). The field notes were used to inform an initial categorization to classify participants working independently or together at certain times. The video analysis process was supplemented by a descriptive statistical analysis. As an initial step, we determined periods in which participants were working on the tasks out of the total sessions' recordings time. We excluded the time spent on "nonworking" such as, briefing, asking questions to the trainer during the experiment, and any other type of interruptions. As a result, 457 minutes 15 seconds (i.e., 75%) of the video recordings were used in the following video analysis. The time spent by each pair to complete the given four tasks varied between 23 to 41 minutes with a mean duration of 25.4 minutes (SD = 1.5 minutes). The time during which participants discussed the tasks before the experiment was not included in the evaluation coding of the sessions' video recordings. However, we consider it as an indication for participants' collaboration building relations. We expected some discussion between the partners before the task execution, so we started the analysis from the point when the small talk was left and the participants started to work with the artifacts (Jakobsen and Hornbæk, 2014).

*Observing participants' behavior:* The participant's verbal communication and visual attention were observed to measure participants' collaboration level and their ability to build a collaborative relation.

*Observing participants' physical arrangement*: The participants' physical arrangement relative to the tiled-wall display and to their partner was observed to describe the different standing position patterns. This helps in analyzing the use of the working space, as it might affect participants' interaction and relation building (Heinrich, 2014).

*Analyzing the collected feedback:* We analyzed the collected participants' answers, opinions, and feedback using the closed-ended and the open-ended questionnaires.

## 6.    Results and Findings

In this section, we describe the results and findings of our conducted user study. We adopted the categorization scheme proposed by Jakobsen and Hornbæk (2014) to describe the participants' behaviors around the collaborative setup. Sessions were divided into two categories, *mutually exclusive* and *exhaustive* states, which characterized whether participants were talking together or not.

We classify the verbal communication as following:

- *Both are talking:* User-A and User-B are both talking typically taking turns. Partners are either engaged in conversation or one speaks and the other is listening to the partner and confirming their paying attention by making brief verbal acknowledgements (e.g., Okay).
- *One is talking:* User-A is talking while User-B is silent and at the same time User-B is not listening or acknowledging their attention to the partner.
- *Silence case:* Both are silent.

We classify the visual communication as the following:

- Both are looking at the display.
- Both are looking at their tablets.
- Both are looking at each other.
- One is looking at the tablet device and the other partner is looking at the display.
- One is looking at the tablet device and the other partner is looking at her/him.
- One is looking at the tiled-wall display and the other partner is looking at her/him.

To study the physical orientation, we analyze several standing positions that were adopted by the pairs during task execution referring to them as **C**ollaborative **P**hysical **P**ositions (**CPP**s). We summarize them into six different possible positions, as later shown in Figure 5, distinguishing these positions based on two main factors, the *physical distance* between the partners and their *view angles*. It is crucial to mention that pairs had the freedom to decide on and change their positions during the experiment. We did not inform them on how to align themselves towards the tiled-wall display or to each other. We observed their taken positions during the analysis and related it with their verbal and visual communication.

Finally, we measure the *collaborative coupling ratio* as the combination of verbal communications and visual communications. Based on this, we classify pairs into two groups: the *loosely coupled* group and the *tightly coupled* group. Further, we also observe the effect of physical arrangements on the partners from the perspectives of verbal and visual communications.

6.1 Collaboration Coupling

*6.1.1 Verbal Communication*

The verbal communication codes of the participated pairs are summarized in Figure 3. Overall, pairs talked approx. 49% of the time and were silent approx. 51% of the time. The silent periods and the talking periods alternated often as is indicated by the mean duration of: both are talking, one is talking, and both are silent (i.e., M = 11.71s, M = 7.77s, and M = 10.86s, respectively). There were many shorter periods of time when only one partner in the pair was talking. The maximum duration when only one partner was talking was 62s while the conversational and silent periods were noticeably longer, reaching to even 280s and 218s respectively. Our video recordings show that pairs frequently talked together (i.e., engaged in a conversation) during their work. We observed that pairs talked together approx. 43% of the time, while the periods of only one partner talking reached approx. 6% of the time.

These results indicate that most of the times participants either talked together or kept silent. This is attributed to the environment, as participants had more than one workspace and it was hard for them to attend the task when only one of them was talking. We found out from the feedback of the open-ended questions that the main motivation for talking during the experiment was to update their partner. The analysis of the verbal communication revealed that participants commented on their work, coordinated the work, suggested solutions, helped each other, and gave directions by talking together, e.g., *"I want to check the lower component of the 3D view, can you please rotate it"* or *"I'm rotating the 3D view. Are you trying to translate it?"*. We observed that many participants tended to verbalize their thoughts during the work. Many of them also suggested that group members need to coordinate their work while working around the wall-display so they can achieve their goals.

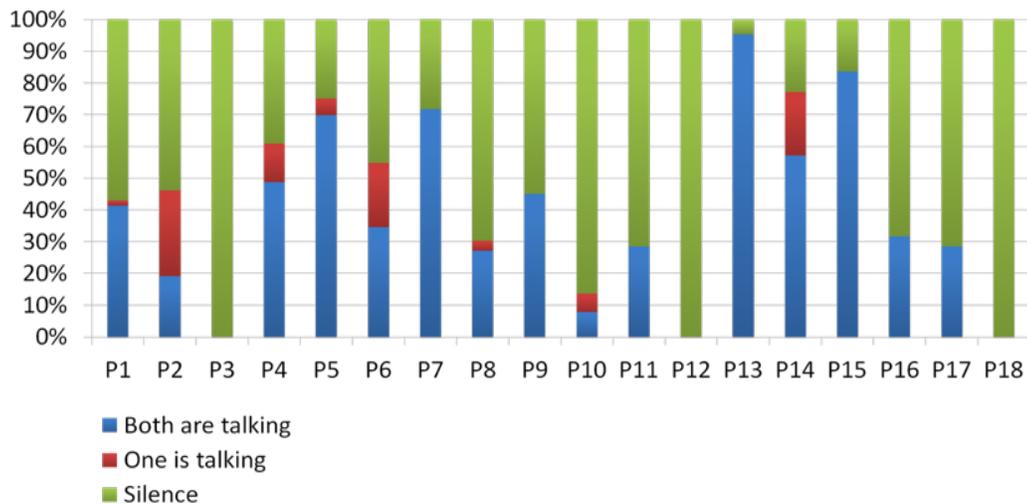

**Fig. 3** The relative amount of time participants in each pair (P1 – P18) spent in different modes of the verbal communication.

Similar to the finding of Russell et al. (2002), we found out that participants talked more often when they did joint work, which helped them to coordinate their actions.

We observed that participants were often more silent when they worked in parallel. Overall, we found out that participants switched frequently and fluidly between periods of being silent and talking. This is also indicated by the short mean duration of the verbal communication codes. Accordingly, periods of silence and with one person talking might indicate *loosely coupled* work, whereas periods where both are talking might typically indicate *tightly coupled* work (Jakobsen and Hornbæk, 2014). However, to better understand coupling in participants' work, we related the verbal communication, the visual attention, and the physical arrangement data (Jakobsen and Hornbæk, 2014), as it is shown in Section 6.1.3.

*6.1.2 Visual Attention*

The visual attention codes of participants are summarized in Figure 4. Overall, pairs spent most of the time (i.e., approx. 76% of the time) looking predominantly at the same device (e.g., the tiled-wall display or the tablet). They continued to look at the same device for relatively long periods at a time (i.e., the display and the tablet, M = 12.86 seconds and M = 13.62 seconds respectively). On the other hand, approx. 24% of the time pairs were predominantly looking at different devices (i.e., the display and the tablet) or their visual attention was divided. Pairs spent much less time looking at each other (i.e., approx. 6% of the time) and did so for much shorter periods at a time (i.e., M = 5.12 sec) compared to when they were looking at the same device. This is not surprising because it becomes more difficult for users to manage three spaces in our setup (i.e., the display, the tablet, and the partner); thus, causing a decrease of time spent by partners of the pairs looking at each other. However, partners in pairs combined different kinds of interactions to overcome this situation such as talking together and glancing to each other from time to time. The summary of the observed visual attention codes suggests that participants spent more time working with the same device (i.e., the display or the tablet: approx. 76% of the time) than working with different devices (i.e., approx. 24% of the time).

In order to have a closer look at how participants visually used the different work spaces, we combined the artifacts' visual states together into one visual direction: *focus to the artifacts* (i.e., both participants are looking towards the display, both participants are looking towards the tablets, or one partner is looking towards the display while the other partner is looking towards the tablet), or *focus to the partner* (i.e., one participant is looking at the partner who is looking towards the artifacts, or both partners are looking at each other). Participants spent approx. 87% of the time *focusing on the artifacts*, while they spent only approx. 13% of the time *focusing on their partner*.

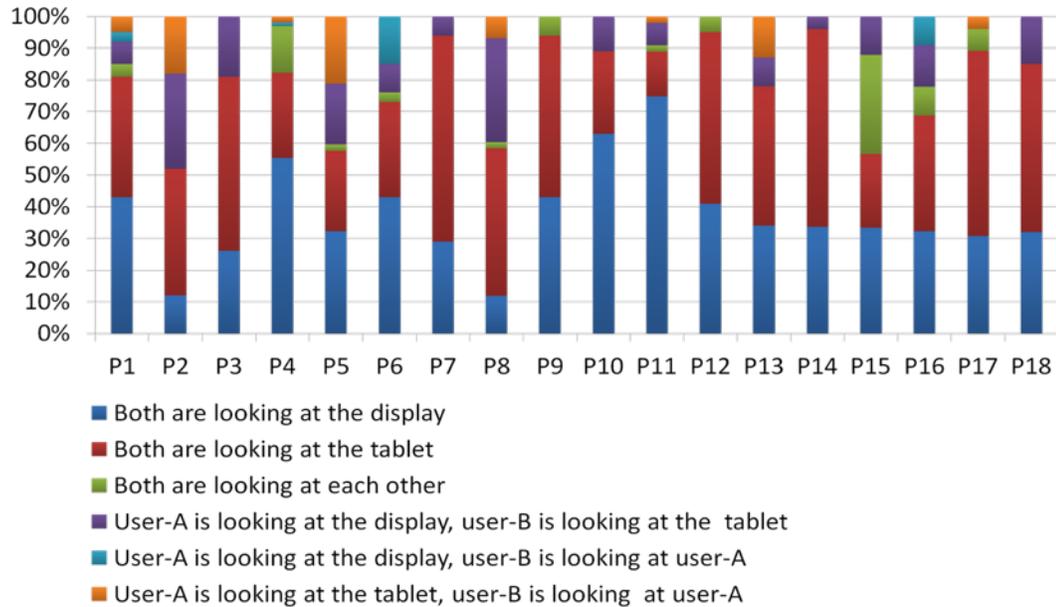

**Fig. 4** The relative amount of time participants in each pair (P1 – P18) spent in different modes of the visual communication.

We noticed that participants made glances at each other for short occasional periods. This behavior was not explained before; because only periods with a predominant gaze pattern lasting more than 5 seconds were coded by Jakobsen and Hornbæk (2014). However, these short periods of glances are important for the partners in maintaining understandability (e.g., switching to joint work or coordinating in the current task) (Jakobsen and Hornbæk, 2014). Sometimes, it was not possible for the partners to keep looking at each other for longer periods especially, when they were working with the tiled-wall display. Thus, they used glances more often to maintain the collaboration. We found that approx. 78% of the participants used glances while working around the tiled-wall display.

*6.1.3 Verbal Communication Combined with Visual Attention*

The partners' verbal communication and visual attention codes can be combined to better understand coupling in their work, as suggested by Jakobsen and Hornbæk (2014). Across sessions, we observed associations between verbal communication and the visual attention codes e.g. in many cases when partners looked at each other (i.e., more than 5 seconds). This indicates that the visual communication in many cases triggers the start of verbal communication, as the partners then expect to receive an explanation of recent occurrences or want to discuss the current/future action. This also indicates that *glancing* is used by partners just to see what the other partner is doing or to acknowledge them. This is the reason to not have any verbal communication or very short verbal communication (like *yes* or *OK*) during glancing. We also observed that partners talked most often when one partner was looking at one of the artifacts and the other partner was looking towards her/him (approx. 75% of the time). This is not surprising, because most often the partner who was looking at the artifact favored verbalizing his/her thoughts and comments on the interaction result reflected on the display without looking at the other partner.

Furthermore, we observed that when both partners were working with the artifacts (i.e., the tablet or the wall-display), then most of the time they were silent. However, they switched from silent to talking mode in between, e.g., when both were working on the same kind of artifact (i.e., 45% with the wall-display and 32% with the tablet) and 21% when they were working on differing artifacts (i.e., one was working on a tablet while the other was working on the wall-display). We found out that in the scenario when both were working on some artifact, the main reasons for talking were building a shared understanding, discussing the possible solution, or coordinating their actions.

*6.1.4 Coupling among Pairs*

In general, results (e.g., see Figure 3 and Figure 4) indicate that pairs used different collaboration styles. We observed that different pairs used different forms of verbal communication and visual attention in order to collaborate around the tiled-wall display. For instance, the partners in pair P14 did not look at each other (see Figure 4), which might indicate *loosely coupled* work. However, they spent most of the time in conversation (i.e., approx. 58%), which can be an indicator for *tightly coupled* work. Another pair (i.e., P8) spent most of the time looking at different artifacts (i.e., approx. 33%, see Figure 4), which might indicate *loosely coupled* work. However, they were talking approx. 28% of the time, which can be indicated as *tightly coupled* work. In such cases, we would consider them as a collaborative couple due to their talking factor because they were in a mixed status most of the time due to the differences in used devices in this test. Many times, partners in the pairs used glancing to communicate with their partners, while in many other occasions they preferred a mixture between talking and looking at each other. However, some pairs worked in silence the entire time with talking periods equal to 0 seconds such as P3 and P18 (see Figure 3), while few pairs (e.g., P13) talked most of the time with talking periods reaching up to 280 seconds. We observed that glancing was used to show a positive action between the partners in some cases (e.g., when a task is completed). However, sometimes it was also used to indicate a few negative actions (e.g., misunderstanding between the partners, less coordination between them, etc.). In both cases, glancing indicates that partners are trying to communicate with each other. Jakobsen and Hornbæk (2014) emphasized that care must be taken in interpreting the coding for a single dimension, based on the differences in the collaboration styles.

Correlating the participants' communication behaviors based on each task is shown in Figure 5. Task 1.1, Task 1.2, and Task 2 were the most collaborative one's due to the mutual interest between the pairs, as the partners in these cases worked on the same view with different interaction techniques. This requires a high rate of synchronization and communication between the partners in order to avoid any conflict. Interestingly, Task 4 was the least collaborative one; although we expected a higher level of collaboration, as the partners were working on the same visual representation in the used application.

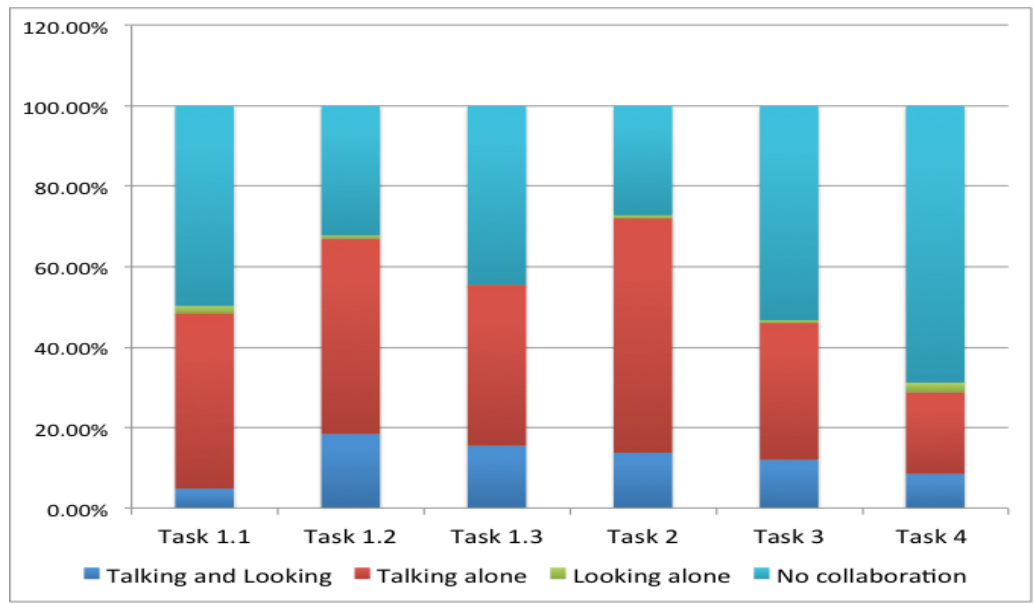

**Fig. 5** The collaboration coupling metrics per task.

When we took a separate look at pairs working on their last tasks, we observed a more implicit collaboration between the partners (i.e., they needed less verbal or visual communication). One reason could be that after working together for some time, they had already developed some harmony between them, which helped them to build implicit collaboration rather than a more explicit collaboration. This indicates that with the passage of time, partners in such collaborative environments may show more implicit forms of collaboration, due to the fact that they may already know their partner's working behavior well as well as utilizing their learning experience of synchronization.

## 6.2 Collaborative Physical Position (CPP) Patterns

The used environment setup in our study seemed flexible in supporting different types of physical arrangements. We observed that pairs took different positions to interact with the wall-display and with each other. However, people's physical positions were also influenced by culture, physiology, and other factors (Jakobsen and Hornbæk, 2014). Previous research (e.g., Heinrich et al., 2014) has described how users' physical arrangements around collaborative displays support the collaboration differently. In our case, we observed the physical arrangement of the body position (i.e., the body and the head orientation of partners towards each other and towards the wall-display) and their standing position in the environment to find out different collaborative physical position (CPP) patterns. This could help in understanding the standing positions' role in collaboration between the partners.

In total, we observed six different CPP patterns (see Figure 6) that were used by the participated pairs in different timings. Out of 18 pairs, 5 pairs used the CPP1 pattern, 4 pairs used the CPP2 pattern, one pair used the CPP3 pattern, 5 pairs used the CPP4 pattern, 2 pairs used the CPP5 pattern, and one pair used the CPP6 pattern. Most of the pairs used only one pattern; however, a few pairs changed their positions slightly but retook their preferred position eventually. Therefore, we counted them in their dominated used pattern.

From the pairs' selection of CPP patterns, we can easily see that the most preferable collaborative physical positions were the ones where the pairs were facing the tiled-wall display and standing either close or far away from each other (i.e., CPP1 and CPP2 patterns) or when they were located diagonally far away from each other (i.e., CPP4 pattern). In the case of the CPP4 pattern, it was easy for the pairs to monitor their partner as well as watch the whole tiled-wall display without turning their neck. Further, we can also see that pairs preferably chose to stand far way, i.e., 4 patterns where they stood far away compared to 2 patterns where they stood nearby and 12 pairs who chose patterns with far away standing positions compared to 6 pairs who chose patterns with nearby standing positions.

This shows that most people even in a collaborative environment, when it is possible, prefer to maintain their own space. The usage of mobile devices as an interaction medium with the shared wall-display enabled them to work in the collaborative environment while keeping their own space. In Figure 7, we show each interaction frequency for all partners, while in Figure 8 we show the average percentage of all interactions lasting at least 5 seconds (not including the *glancing* interaction) in each pattern done by all pairs who used that pattern.

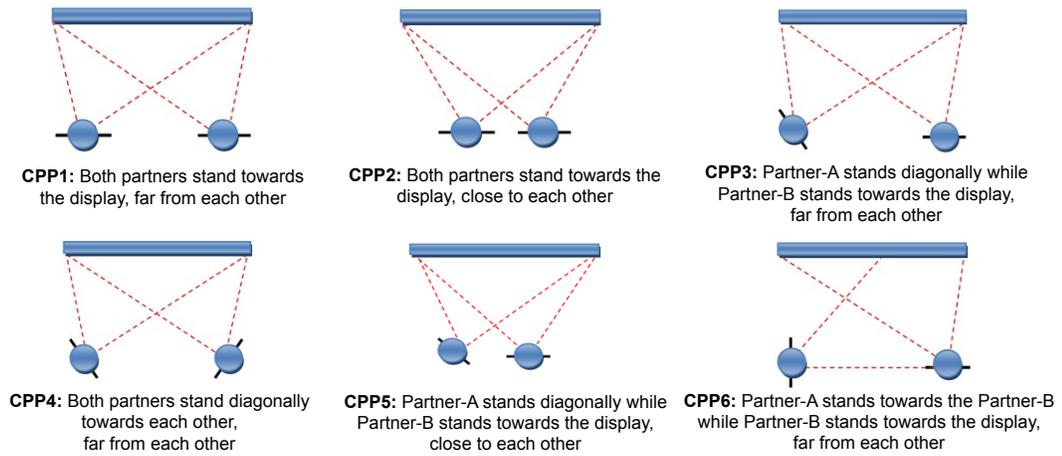

**Fig. 6** The six Collaborating Physical Position (CPP) patterns found during the experiment.

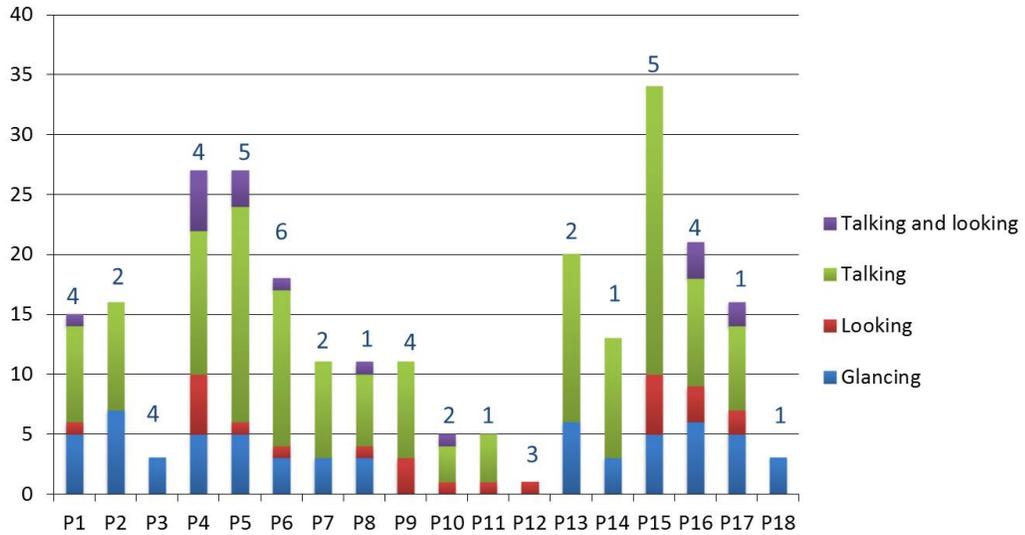

**Fig. 7** Frequency of interactions inside the pairs. Top of each bar shows the pair's used CPP id.

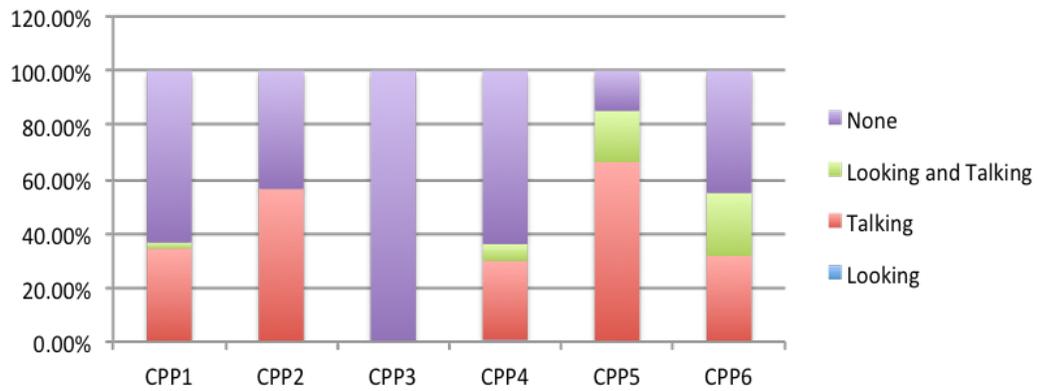

**Fig. 8** Average percentage of all interactions in each collaborative position pattern.

6.3   Collaborative Coupling and Physical Positions

The participants' behavior codes (i.e., the verbal communication and the visual attention) and their physical positions during the collaborative task can be studied together to better understand the coupling level between the partners (Tang et al. 2006; Isenberg et al. 2012; Jakobsen and Hornbæk 2014). In this context, the coupling styles are described as: *tightly coupled* referring to when partners are working together closely by looking at each other and talking together, and *loosely coupled* referring to when partners spent most of their time looking at artifacts without looking at each other or talking together. To better understand the coupling level, we can relate the participants' standing positions to the encodings of the verbal communication and the visual attention, similar to the work of Jakobsen and Hornbæk (2014).

First, verbal communication differed based on the pairs' standing positions, e.g., we observed that pairs were talking more frequently and were less silent when one of them or both of them were standing diagonally towards each other (i.e., CCP4 and CCP5). Second, we observed that pairs were more silent when they were facing the display (i.e., CCP1 and CCP2). We also observed that participants were looking at each other more often when they were standing diagonally compared to when they were facing the display. Pairs who spent most of their time standing diagonally towards each other (such as P4, P5, P15, and P16) had relative high percentages of verbal communication (i.e., 47%, 75%, and 82% respectively) as well as high percentages of visual attention (i.e., 48%, 75%, 82% respectively), see Figure 7. In this standing position, one partner faced the other partner and in nearly the same line of sight faced the display. While standing diagonally, they were able to sense when their partners were looking at them without the need of rotating their heads. This allowed them to focus on the wall-display while being able to track their partners' point of focus. On the other hand, when pairs stood facing the display, they had to turn their heads to look at each other compared to when they stood diagonally.

From these observations, we can conclude that partners preferred to stand diagonally in our setup where they were able to give a high level of mutual attention including conversation. There was only one pair that took in the CCP3 position but they had very limited interactions during the test. As only one pair appeared in this scenario, we could still consider standing diagonally as the optimum standing position, which increased participants' collaboration.

By relating physical positions with the behaviors' codes, we observed that pairs were often more tightly coupled when they were physically standing towards each other and often more loosely coupled (i.e., they were more silent and looked more towards the artifacts) when they were physically standing towards the display. However, we found that pairs shifted many times from loosely coupled to tightly coupled conditions and vice versa through physical movements. We also observed that pairs prefer to stand physically close to each other when they want to work together and far from each other when they prefer to work independently around the wall-display, similar to the finding of Tang et al. (2006) and Jakobsen and Hornbæk (2014). Pairs seemed to switch frequently between independent and joint work. This is also indicated by the short mean duration of the codes for different modes of communication, attention, and standing position between the partners.

6.4 Workspace Awareness

*Workspace awareness* is an important factor in collaborative environments that affects coupling levels, which are described as the personal understanding of interacting with the shared workspace (Heinrich et al., 2014). In our setup, we found out that engaging the mobile devices with the shared wall-display environment introduces different workspaces, where the participant's focus is driven to these spaces during the test. We came up with four different workspaces (see Figure 9) in our setup after analyzing the results:

- **Display Workspace (DW):** It is defined as the space partners look at if they want to work with the display (i.e., both partners are looking at the shared-wall display).
- **Mobile Workspace (MW):** It is defined as the space partners look at if they want to work with the mobile device; here both are looking at their tablets.
- **Multi-Devices Workspace (MDW):** It is defined as the space when one of the partners is looking at the display, while the other is looking at the mobile device.
- **Collaborative Workspace (CW):** It is defined as the space when the partners *engage* in a *collaborative relationship*. Therefore, it represents the workspace where there is a direct interaction between the partners, verbally or visually.

All participants were observed in one of these four mentioned workspaces. We did not notice participants deviating from these mentioned workspaces, thus we do not include the state "elsewhere".

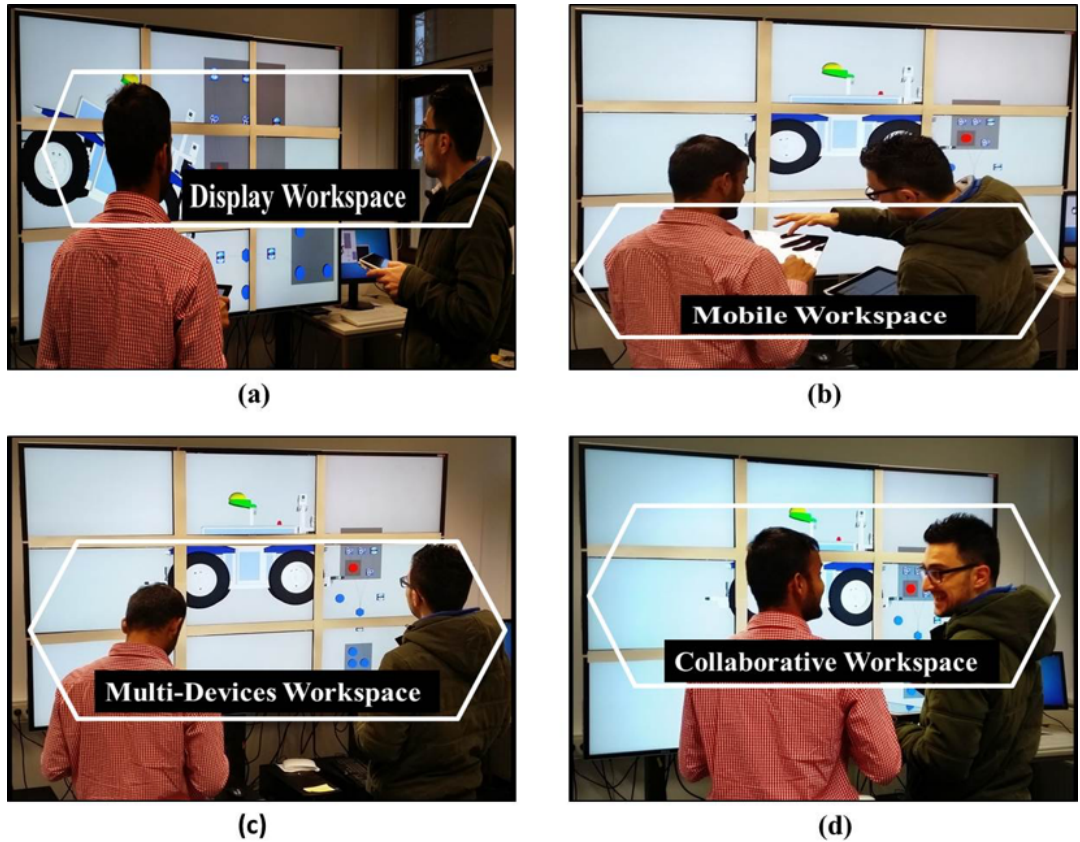

**Fig. 9** States of view when working around the wall-display using tablets.

From the description of the results and the findings, we observed that participants were in the *MDW* state (i.e., frequent transition between the used devices) most of the times, according to the situation in the executing task. This indicates that the shifting of their focus was motivated by the used device (whether the mobile device, the wall-display, or both), which helped them in solving their current task, as well as to coordinate with the current mode of their partner. In tasks requiring a lower level of coordination, the partners focused more on either the *display workspace* or the *mobile workspace*. However, we observed that even in these two workspaces, the partners maintained a sense of the whereabouts and activities of the other partner. For example, in one task where partners had to work together to find out the shortest failure path between the two specified components in the system, it was necessary for them to communicate with each other at the beginning of this task in order to synchronize their activities. Therefore, they started in the *collaborative workspace* mode. After the initial conversation with their partner, most of the pairs then started observing the shared wall-display, which means they entered the *display workspace*. However, after a short time the partners came to an agreement between them through verbal and visual messages to arrange further steps between them. They kept an eye on what the other partner was doing through *glancing* and then conferred, commented, coordinated, or gave feedback to the partner by verbal communications (e.g., by saying *"I am expanding the required node."* or *"I found the required node."*). These kinds of messages enabled them to continue working individually on their own mobile devices. However, they were trying to synchronize their activities to execute the underlying task by shifting their focus simultaneously between the shared wall-display and their own mobile devices. This behavior was also frequently observed in many other collaborative tasks.

Overall, most of the participants were coordinating with their partners and continuously shifting their focus between the *display workspace* and the *mobile workspace* while executing their tasks. For example, one participant in the *display workspace* mode said *"I want to check the lower component of the 3D view, can you rotate it please?"*, and the partner replied *"I will translate the 3D view first, then it will be your turn"* while shifting his focus to the *mobile workspace* mode accordingly. They also monitored each other's work by saying *"I am rotating the view, so can you try to translate it!"*. This kind of communication between them was helpful in solving some conflicting cases that might occur. In many cases, participants were helping their partners with verbal and visual communication where it was needed. Other verbal communication messages, such as *"go to the next task"* or *"did you move the 3D view up?"* were used frequently amongst the partners to successfully execute the collaborative tasks.

In our study, we collected this data and analyzed it by studying the recorded video and audio material carefully. We observed that partners were often talking and/or using glancing during individual work. This made it easier for them to operate and to understand the required task and the required steps to solve the underlying tasks. However, participants shifted to the collaborative workspace mode not only to collaborate but also to monitor their partner's work in order to maintain awareness of what the partner was doing and how to react accordingly. In fact, in our study 39% of the participants agreed that they monitored their partner's work during the test. This is also indicated by analyzing the overall test duration of the used codes for different modes of verbal communication, visual attention, and glances. Overall, verbal communication was used more often between the partners compared to visual communication. As a conclusion, they were able to move between the individual working modes (i.e., the display workspace, the mobile workspace, and the multi-devices workspace) and the group working mode (i.e., the collaborative work space).

In general, the ability of participants to interact with the shared wall-display using multiple mobile devices allowed the partners to simultaneously work in different modes.

## 6.5 Subjective Feedback

At the end of the test, we collected considerable qualitative feedback from the participants through the closed-ended (see Figure 10) and the open-ended questionnaires. Overall, participants agreed that our built collaborative setup is good for collaborative work. Most of the participants (i.e., approx. 85%) agreed that they felt satisfied and comfortable while working in our collaborative setup. They responded that such a collaborative setup provides people in a team the flexibility to better understand each other and to find solutions for complex tasks. They also responded positively towards using the mobile devices (tablets in our case) for interacting with the shared tiled-wall display, as this enabled them to move freely in the working space. Around 83% of the participants felt satisfied using the mobile devices in such a collaborative setup, as it was easy for them to synchronize their eyes between the different devices (i.e., the tablet and the display) and their partner.

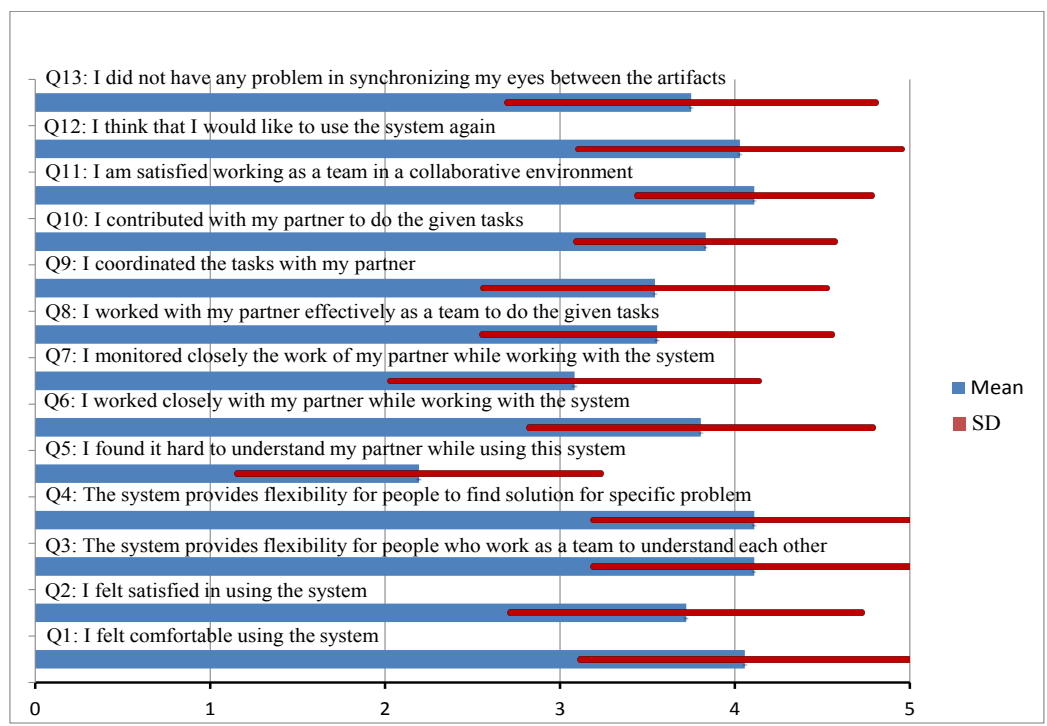

**Fig.10** Closed-ended questionnaire results showing participants' feedback after the system evaluation.

However, some of the participants commented in the open-ended feedback that they had difficulties at the beginning to cope with the environment, but eventually were able to adjust themselves to the environment after only a short period of time. Further, many participants admired that the environment supports users' communication and interactions, which provides a good solution to enhance team work. For example, one participant commented that: *"the environment enhances the process of finding solutions for specific problems, the use of the tiled-wall display is a good solution to view the partner's actions and your actions in a single scene. At the same time, the usage of a tablet for each participant makes it easier for both partners to interact with the tiled-wall display and with each other"*. A few participants commented that partners need to collaborate with each other to reach a consensus; otherwise, they will not be able to find a solution that would satisfy all group members. Many of the participants agreed that continuous interaction between the partners is needed to complete their tasks effectively and efficiently, while few participants commented that the interaction between the partners is only needed in some situations. They commented that interaction is essential when partners work jointly but is not required when working in parallel.

### 7. Conclusion and Future Work

In this paper, we reported the results of a controlled user study that was conducted to measure the *collaborative coupling factors* between partners who worked in a collaborative setup, built with a shared tiled-wall display and multiple mobile devices for interaction. We measured the verbal and the visual communication between the partners. The verbal and visual communication codes examined the extent of participant coupling while working during the test. We found variations in how pairs collaborated around our collaborative environment setup and that partners talked more often when they performed joint work activities in order to coordinate their actions. Furthermore, they were observed to be silent more often while performing individual work activities. We combined the two codes to better understand coupling in partners' work. The clearest association was found to occur between partners looking at each other and talking. Overall, participants seemed to switch frequently between periods of silence and talking. This was also indicated by the short mean duration of the verbal communication codes. The tiled-wall display seemed flexible in supporting different simultaneous interactions by the partners. The usage of mobile devices allowed partners to work together with less risk of distraction, as each partner had her/his own mobile device. Further, these mobile devices helped the partners to interact with the shared-wall display and work in parallel, as well as to interact with each other.

Further, we also found collaborative physical position (CPP) patterns frequently used by the pairs. We noted six CCP patterns and observed some interesting findings, e.g.: when equipped with mobile devices for interaction with the wall-display, partners prefer to have their own personal space except when they engage in more conversational interaction; more frequent silence periods and less talking periods were observed when partners were facing the display; standing far away while diagonally facing their partner and the wall-display provides the most relaxing position and enables the pairs to build effective collaborative relations through all means of interaction modes between the pair of partners. We observed that when standing diagonally, partners could clearly perceive the content on the wall-display and their visual field covered sufficient space of the display as well as their partner. We found out that partners were more often tightly coupled when they were physically standing facing each other and more often loosely coupled when they were physically facing the display. However, some pairs were shifting from loosely coupled to tightly coupled through physical movement and vice versa. In general, we did not find any significant differences in the collaboration ratio based on gender, culture, or age.

We conclude that for a proper collaborative work in our collaborative environment, three main points should be considered: First, both partners need to be aware of the importance of collaboration during the task execution in such a collaborative environment. Second, both partners need to communicate between themselves regarding the targeted task. Third, partners need to position themselves to facilitate communication. We suggest to encourage discussions between the partners before starting the work in the collaborative environment, as even a short discussion phase would be useful for establishing an early relation between the partners. This would help them to become aware of the importance of collaboration, which is useful in the later phase during which they must find the solution together.

This study can be used as an initiative study to encourage researchers from different domains to seriously consider studying the usage of mobile devices in enhancing the collaboration between people in computer-supported collaborative setups, which would open new doors to utilize the power of these smart mobile devices. However, we did not correlate between the collaborative coupling and the pairs' effectiveness or efficiency in task completion. We believe that if they were compared with each other, the traditional *tightly coupled* factor might not indicate a high performance. This is because we observed that some partners were moving freely in the environment and did not talk or look at each other most of the time. However, they were able to achieve their tasks effectively and efficiently. This would be the next step in our research, as we believe that in such environments (that utilize the power of smart mobile devices) a new definition for *effective collaborative coupling* is needed. Moreover, we intend to perform further studies in order to specify the definition of the coupling factor in mobile devices-supported collaborative setups. Furthermore, we plan to apply the concept in some real scenarios and will ask the real-end users to apply the setup in their work, as well as evaluating it from these real-end users' perspectives.